# Correlations among magnetic, magnetocaloric and magneto-transport properties in HoNiSi

Sachin Gupta and K. G. Suresh*

Department of Physics, Indian Institute of Technology Bombay, Mumbai-400076, India

## Abstract

Magnetic, magnetocaloric and magneto-transport properties of polycrystalline HoNiSi have been studied. The compound crystallizes in orthorhombic crystal structure and orders antiferromagnetically at $T_N$=4.6 K. Magnetization isotherms show curvature at small fields, revealing a field induced metamagnetic transition. Magnetocalroic effect (MCE) has been estimated using magnetization data and is found to be 12.8 J/kg K for a field of 50 kOe. Application of field reduces the resistivity near $T_N$, which results in a large negative MR. The maximum value of MR has been found to be -24% at 4 K for 50 kOe. Below 4 K, the compound shows positive MR with shape changing with field.

*Corresponding author email: suresh@phy.iitb.ac.in

## 1. Introduction

RTX (R=Rare earth element, T= transition metal element, and X is p-block element) is a very huge reservoir of interesting magnetic compounds in the rare earth based intermetallics family. Many of its members show versatile magnetic and electrical properties in a wide range of temperatures. A review of these properties of a large number of members of this series has been published by us recently.[1] It has been observed that some RTX compounds show large magnetocaloric effect (MCE)[2-8] and magnetoresistance (MR)[5,9,10] around their ordering temperatures. The large MCE in these compounds arises due to large magnetic moments and large slope change in magnetization at different temperatures around the ordering temperatures. The field induced metamagnetic transition in antiferromagnetic compounds also results in large MCE and MR. It is well known that the both MCE and MR are dependent on the coupling of magnetic sub-lattice with the magnetic field and therefore, these two properties are also related. In fact, such a correlation has been shown in many materials.[5,11,12]

In RNiSi series belonging to the RTX family, the compounds with R=La-Nd crystallize in LaPtSi-type tetragonal structure.[13,14] LaNiSi was found to show superconductivity with a $T_C$ of 1.2 K.[13] CeNiSi shows Kondo behavior, while NdNiSi shows two antiferromagnetic transitions, at $T_{N1}$ = 6.8 K and $T_{N2}$ = 2.8 K.[14] The authors also reported the effects of hydrogenation of these compounds and observed that on insertion of hydrogen, the Kondo effect in CeNiSi strengthens, while the antiferromagnetic transitions in NdNiSi disappears and does not show any magnetic ordering down to 2 K in hydrogenated NdNiSi. Szytuła et al.[15] have reported magnetometric and neutron diffraction data for RNiSi (R=Tb-Er) compounds. They observed that these compounds crystallize in TiNiSi-type orthorhombic structure with space group Pnma and show antiferromagnetic ordering at low temperatures. Neutron diffraction studies show square-modulated magnetic structure in TbNiSi and DyNiSi at 1.5 K, which transforms to sine-modulated structure at 9 and 3 K, respectively and is stable up to their $T_N$s. HoNiSi and ErNiSi show collinear magnetic structure at 1.5 K, which transforms into sine-modulated magnetic structure at 3.2 and 2.2 K, respectively. Recently, in the search of large MCE and MR, we have studied ErNiSi of this series. This compound shows large MCE and MR around its ordering temperature.[5] Motivated by the results of ErNiSi, we tried HoNiSi, which shows good MCE and MR. We find that though MCE behavior has some similarities, the MR behavior of HoNiSi is different from that of ErNiSi. In this paper, we report magnetic, heat capacity, magnetocaloric, transport and magneto-transport properties of HoNiSi.

## 2. Experimental details

Polycrystalline sample of HoNiSi was prepared by arc melting of its constituent elements taken in stoichiometric amount in water-cooled copper hearth in presence of argon atmosphere. The formed ingot was flipped up and melted several times for better mixing of constituent elements. After melting the sample was sealed in evacuated quartz tube and annealed for 100 h at 800 ºC. To check the phase purity the x-ray diffraction pattern was obtained from X'Pert PRO diffractometer using Cu Kα radiation. The Rietveld refinement shows TiNiSi-type orthorhombic crystal structure with space group Pnma. The lattice parameters obtained from the refinement are close to reported values. The magnetization, M(T) and M(H), and the heat capacity measurements were carried out on quantum design Physical Property Measurement System (PPMS) model-6500. The resistivity measurements, with and without field, were carried out on home made system along with Oxford 8 T superconducting magnet using standard four probe method in longitudinal geometry in a home made system.

## 3. Results and discussion

The magnetic susceptibility ($\chi$) for the field of 500 Oe and in the temperature range 1.8-300 K (left-hand axis) is shown in Fig. 1 along with Curie-Weiss fit (right-hand axis). The cusp-like behavior of $\chi$ at low temperatures indicates the antiferromagnetic ordering in this compound. The inverse susceptibility data was fitted by modified Curie-Weiss law, $(\chi-\chi_0)^{-1} = (T-\theta_p)/C_m$, where $\chi_0 = 0.00147$ emu/mol Oe is the temperature independent susceptibility, arising due to the diamagnetic and Van-Vleck contributions, $C_m$ is the molecular Curie constant and $\theta_p$ is the paramagnetic Curie temperature. The value of effective magnetic moment ($\mu_{eff}$) and $\theta_p$ estimated from the modified Curie-Weiss law fit are 10.7 $\mu_B$/Ho$^{3+}$ and -7.5 K, respectively. The negative $\theta_p$ confirms antiferromagnetic ordering in this compound.

The field dependence of magnetization for HoNiSi is shown in Fig. 2 for different temperatures with an interval of 2 K. At low fields, the magnetization increases linearly with field and also with temperature (i.e., the magnetization at 4 K is higher than that of 2 K) below $T_N$. These facts again confirm the antiferromagnetic state of the compound at low fields and low temperatures. However at higher field the trend changes, the magnetization shows curvature and decreases with increase in temperature (i.e., the magnetization at 4 K is lower than that of 2 K). These facts indicate a field-induced metamagnetic transition in this compound. Above $T_N$, the magnetization is almost linear, typical of paramagnetic materials.

Fig. 3 shows the zero field heat capacity data, which shows a λ-shaped peak at 4 K, which is close to the antiferromagnetic transition temperature obtained from the magnetic susceptibility data and arises due to the onset of antiferromagnetic ordering. One can note that the heat capacity shows a hump above the ordering temperature, which might be due to the Schottky anomaly. A similar anomaly is also seen in some other Ho compounds,[16] which arises due to the crystal field split ground state doublet in the Kramer Ho ion.

The MCE in this compound has been estimated from magnetization data using Maxwell's relation, $\Delta S_M = \int_0^H \left(\frac{\partial M}{\partial T}\right)_H dH$. The temperature dependence of isothermal magnetic entropy change ($\Delta S_M$) in different fields is shown in Fig. 4. The compound shows large $\Delta S_M$ around its ordering temperature, with a value of 12.8 J/kg K for the field of 50 kOe. The $\Delta S_M$ is comparable to that of many potential refrigerant materials such as HoRhSn (13.2 J/kg K, $T_N$=6.2 K ),[17] HoRhGe (11.1 J/kg K, $T_N$=5.5 K),[16] HoMnO$_3$ (13.1 J/kg K for 70 kOe, $T_{N,Ho}$ =4.6 K),[18] Er$_2$Ni$_2$B$_2$C (9.8 J/kg K, $T_N$=6 K),[19] all subjected to 50 kOe. The refrigerant capacity (RC) estimated by, RC= - $\Delta S_M^{max} \times \delta_{FWHM}$, ($\delta_{FWHM}$ is the full width at half maximum) is found to be 312 J/kg for the field of 50 kOe. This value is comparable to that of many intermetallic compounds reported earlier.[1] Large $\Delta S_M$ and RC values in this compound are attributed to the metamagnetic transition.

To know more about the magnetic state of the compound, we have fitted -$\Delta S_M$ vs. T and -$\Delta S_M$ vs. H plots at different fields and temperatures, using the relation, -$\Delta S_M$=C$_m$H$^2$/2T$^2$, where H is the applied field. The C$_m$ estimated from the fit is found to be 15.6 emu K/ mol Oe for the field of 10 kOe. This value is very close to that obtained from the Curie-Weiss fit. (14.5 emu K/

mol Oe). From the inset of Fig. 4(b), one can see that the fit is good only at higher temperatures (i.e., 21 and 41 K). This suggest that the spin fluctuations are present and determine the MCE behatvior in this compound.

The zero field electrical resistivity is shown in Fig. 5. The inset shows the low temperature data in different fields. The compound shows an increase in resistivity with temperature, reflecting its metallic behavior. At low temperatures, the resistivity data show a slope change, which arises due to the onset of magnetic ordering. From the inset of Fig. 5, one can see that the resistivity shows a decrease near the ordering temperature on the application of field. On further decrease in temperature, the trend is reversed.

The MR has been estimated from the field dependence of resistivity using the formula MR% = [$\rho(H,T)- \rho(0,T)/ \rho(0,T)$]×100, where $\rho(H,T)$ is the in-field resistivity and $\rho(0,T)$ is the zero field resistivity. Fig. 6 shows the temperature dependence of MR in 20 kOe field. One can see that MR becomes strongly negative with decrease in temperature and shows a maximum value near $T_N$. It then starts decreasing in magnitude with further decrease in temperature. At the lowest temperature, it is positive in sign. The inset in Fig. 6 shows the field dependence of MR at different temperatures. Down to 4 K, the MR value increases with temperature and increases with field. It shows a MR of -24% for the field of 50 kOe. The large negative value of MR in this compound arises due to the metamagnetic transition and is larger than that of most of the *RTX* compounds reported.[1] The MR trend is different below 4 K; at 3 K, it increases with field, shows a positive maximum and then starts decreasing, becoming strongly negative at higher fields. That is the MR at 3 K shows a sign change from positive to negative as the field is varied. On the other hand, at 1.5 K, the MR shows an initial increase with field, reaches a maximum, followed by a saturation trend with positive MR. The negative MR down to 4 K (main panel of Fig. 6) is attributed to the suppression of spin disorder scattering by the applied field. However, at 3 K (below *$T_N$*), the antiferromagnetic coupling is strong and therefore the application of low field enhances the spin fluctuations due to reorientation of spins and results in positive MR. When this field becomes sufficiently large, the MR starts decreasing and becomes negative thereafter. It is evident from the MR at 1.5 K that the field is not enough to break the antiferromagnetic coupling and hence the MR does not become negative even at the highest field.

One can also see that HoNiSi shows considerable negative MR at 20 K, which is well above the ordering temperature. The negative MR in the paramagnetic regime generally arises due to the suppression of spin fluctuations and shows quadratic field dependence. The inset in Fig. 6 shows a good fit to MR, which confirms the quadratic dependence, again highlighting the presence of spin fluctuations (as in the MCE data). As Ho moments are well localized, it is quite reasonable to assume that the spin fluctuations arise from the Ni sublattice.

In comparison, ErNiSi shows negative MR down to 3 K, while in case of HoNiSi the negative MR is seen down to 4 K. Both the compounds show considerable MR at higher temperatures, which is attributed to the suppression of spin fluctuations in the paramagnetic regime. The values of MCE and MR in the case of ErNiSi are higher than those of HoNiSi. It can also be noted from these compounds that metamagnetic effect is weaker in HoNiSi as compared to ErNiSi. Because of it HoNiSi shows very weak tendency of saturation of magnetization at 50 kOe, while in ErNiSi shows almost saturation of magnetization at 50 kOe. This may also be the reason for higher MCE and MR in ErNiSi as compared to HoNiSi.

## 4. Conclusions

HoNiSi crystallizes in the orthorhombic structure and shows antiferromagnetic ordering at about 4 K. On application of field, it undergoes metamagnetic transition, as many other members of *RTX* series. It shows large MCE and MR around $T_N$, which arise due to the metamagnetic transition. The positive MR below 4 K and its field dependence clearly indicate the presence of spin fluctuations. It is also noted that the highest MR value observed in this compound is larger than that of most of *RTX* compounds studied till now. The large MCE is attributed to the metamagnetic transition, while its field dependence reveals the role of spin fluctuations.

## Acknowledgments

SG thanks CSIR, New Delhi for granting senior research fellowship and UGC-DAE Consortium for Scientific Research, Indore for providing magneto-transport facility, accommodation and travel support. The authors thank Dr. R. Rawat, UGC-DAE CSR for fruitful discussions.

**Figure Captions**

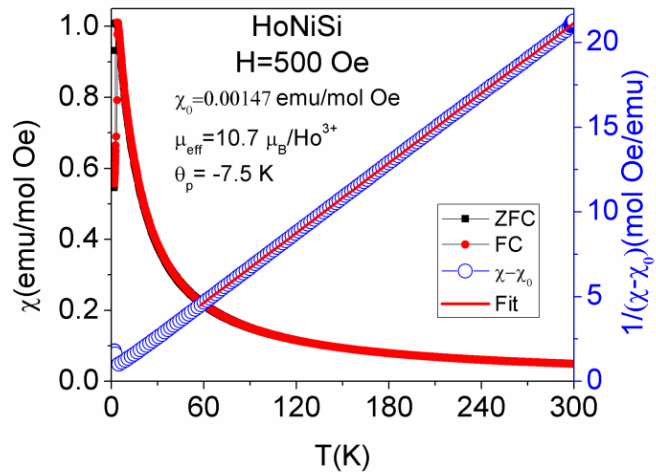

FIG. 1 The temperature dependence of dc magnetic susceptibility in a field of 500 Oe (left-hand axis). The inverse susceptibility along with the modified Curie-Weiss law (right-hand axis).

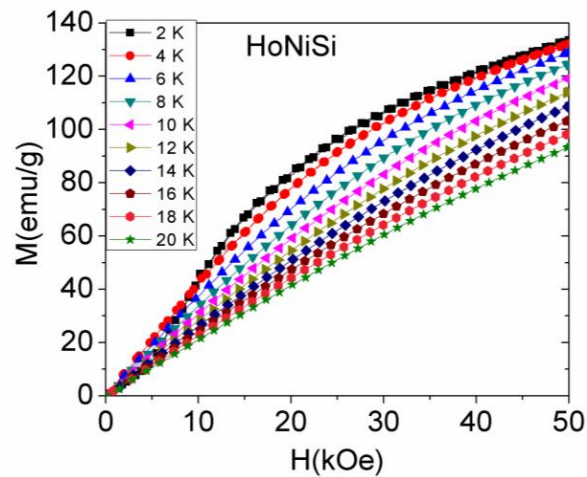

FIG. 2 The field dependence of magnetization (M) at different temperatures.

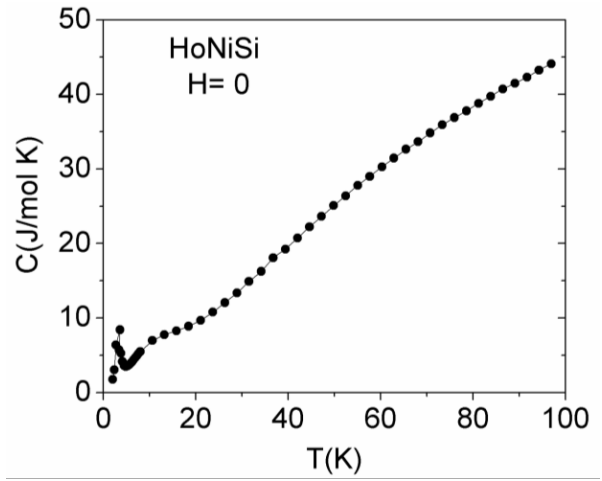

FIG. 3 The temperature dependence of zero field heat capacity.

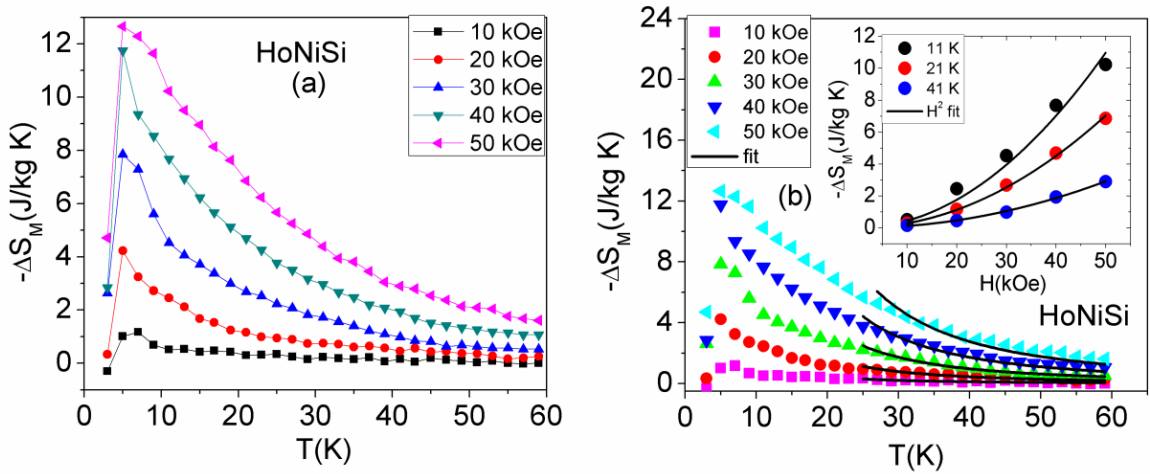

FIG. 4 (a) The temperature dependence of $\Delta S_M$ at different fields for HoNiSi. (b) Fit to - $\Delta S_M$ at different fields (main panel) and temperatures (inset).

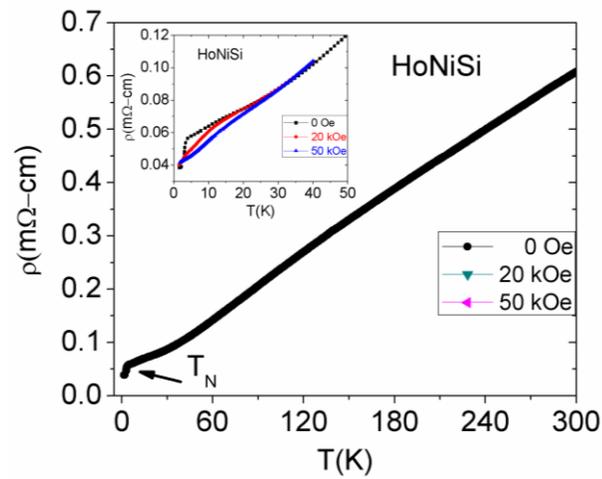

FIG. 5 The temperature dependence of zero field electrical resistivity. The inset shows low temperature data in different fields.

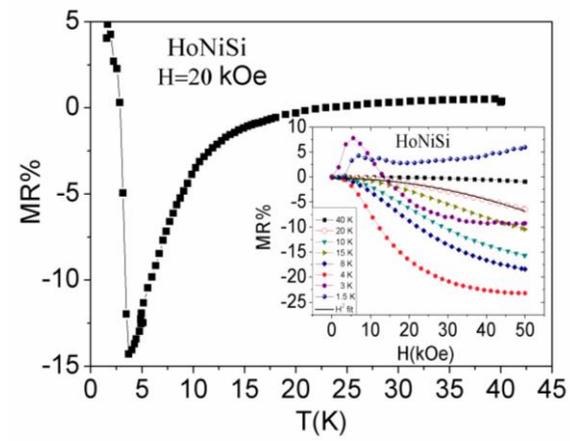

FIG. 6 The temperature dependence of MR in 20 kOe. The inset shows the field dependence of MR at different temperatures along with a $H^2$ fit at 20 K.